# Preserving the topological surface state: FePc on Bi$_2$Se$_3$

*Rejaul Sk[1], Imrankhan Mulani[1], and Aparna Deshpande[1*]*

[1]Scanning Tunneling Microscopy and Atom Manipulation Laboratory, Division of Physics and Center for Energy Science, h-cross, Indian Institute of Science Education and Research (IISER) Pune 411008, India

ABSTRACT. Iron phthalocyanine (FePc) molecules were investigated on the topological insulator surface Bi$_2$Se$_3$ using scanning tunneling microscopy (STM) and spectroscopy (STS) at 77 K in ultra high vacuum (UHV). Sub-molecular resolution STM images provide the atomic scale registry of the adsorption site of the molecules, on-top or bridge, with reference to the top surface Se atoms of the Bi$_2$Se$_3$ surface. STS data reveal a slight shift in the HOMO and LUMO orbitals of the FePc molecules with respect to the specific binding site. Close to Fermi energy the STS measurements show that the Dirac point of Bi$_2$Se$_3$ remains unchanged after depositing FePc molecules. This is indicative of the protection of the topological insulator surface state (TISS). We speculate that though the Fe atom has magnetic properties, the ligands attached to Fe in the Pc ring influence the magnetic behavior of the molecule upon adsorption on the $Bi_2Se_3$ substrate. Our work puts forth an example to advance the strategic platform for tuning the TISS interfaces with different molecules and for designing surfaces and interfaces with time reversal symmetry protected magnetic textures.

INTRODUCTION

For about a decade now, topological insulators (TIs) have been at the forefront of condensed matter physics research. TIs possess strong spin-orbit coupling and are characterized by a distinct band structure showing bulk band gap and conducting surface and edge states[1]. The topological insulator surface states (TISS) characterized by Dirac points, are gapless and exhibit time reversal symmetry. The electrons in these surface states of topological insulators are endowed with intriguing properties like the coupling of the electron spin to its momentum giving rise to spin-momentum locking[2]. This leads to a suppression of the elastic backscattering of electrons and no spin flip can occur. A vital consequence of this stoppage of spin flip means that the spins of the electrons can flow unhindered and constitute spin currents. This fundamental phenomenon thus makes the TIs viable for spintronics. Bismuth selenide, $Bi_2Se_3$, belongs to the TI family. The atoms, Bi and Se, are stacked in the form of quintuple layers with Se layer being the topmost layer[3]. The TISS in $Bi_2Se_3$ can also contribute to generating spin orbit torque (SOT) in magnetically doped heterostructures where the torque from this TISS has been shown to exceed the one in heavy metal ferromagnetic heterostructures[4]. The massive SOT from the Cr-doped TI can be controlled by applying an electric field using a top gate fabricated on a Hall bar of the Cr-doped TI[5]. Thus TIs in addition to being the rich powerhouse of fundamental physics have emerged as robust and promising materials for spintronics and quantum computation[6,7].

Quasiparticle scattering phenomena in TISS of $Bi_2Se_3$ off defects probed by STM and STS have revealed the robust symmetry protection for TISS[8]. Doping of TISS with magnetic Fe atoms in the bulk[9] resulted in a band opening in angle-resolved photoemission spectroscopy (ARPES) of ~ 100 meV, and surface doping[10] gave a similar band gap which was attributed to the out of plane anisotropy of the magnetic moments of Fe. However this report was contradicted by another ARPES experiment where Fe deposition was carried out under several different conditions. No band gap opening was observed in this case. Thus the TISS remained protected[11]. Another set of measurements with STM, and x-ray magnetic circular dichroism (XMCD), and density functional theory (DFT) calculations also reported a lack of band gap opening and explained it in terms of an in-plane magnetic anisotropy of Fe atoms. This absence of magnetic order of Fe atoms normal to the $Bi_2Se_3$ surface plane justified their observation[12]. A comprehensive STM, ARPES, and XMCD study of Co atoms on $Bi_2Se_3$ reported an intact Dirac cone and no ferromagnetic ordering[13]. Thus with single magnetic atoms on the $Bi_2Se_3$ surface the results have been quite diverse. A noteworthy STM report showed that Fe doping on $Bi_2Se_3$ shifts the Dirac point but upon introducing a self-assembled organic molecular layer of perylene-3,4,9,10-tetracarboxylic dianhydride (PTCDA) between Fe and $Bi_2Se_3$ the effect of Fe doping is destroyed and the Dirac point returns to its initial position as that for pristine $Bi_2Se_3$[14]. An ARPES and DFT study showed that $Bi_2Se_3$ upon doping with custom synthesized with molecules $C_{60}$, $H_2Pc$ and $H_2Pc^s$ with the goal of enhancing the molecule-TI interaction, the interface properties could be tailored. A shift of the Dirac point and hybrid interface states were reported [15]. A large family of phthalocyanines MnPc, CoPc, CuPc have been studied on the surface of $Bi_2Te_3$- another member of the TI family, where the networks of these molecules and $Bi_2Te_3$ create an interface engineered by charge transfer processes [16,17]. However $Bi_2Se_3$ has not been probed much for such surface-interface interactions. A recent report of CoPc on $Bi_2Se_3$ shows a charge transfer from molecule to substrate[18].

In our work presented here we have chosen the iron phthalocyanine molecule, FePc, to dope the $Bi_2Se_3$ surface. Given the vivid reports of Fe atoms on $Bi_2Se_3$[9-12,14] . We would like to explore the interaction of FePc with $Bi_2Se_3$ and question– does the central Fe atom govern the on-surface behavior of FePc upon adsorption on $Bi_2Se_3$? Do the ligands of Fe participate in its interaction with the $Bi_2Se_3$ surface? How does the FePc adsorption influence the surface state of $Bi_2Se_3$? Seeking answers to these queries forms the premise of our paper.

EXPERIMENTAL METHODS

The experiments were performed using an Omicron ultra-high vacuum (UHV) low temperature (LT) scanning tunneling microscope (STM) at 77 K with a base pressure of $5\times10^{-11}$ mbar. The $Bi_2Se_3$ single crystal was grown using the modified Bridgman technique. First, stoichiometric mixtures of high purity elements Bi, 99.999% and Se, 99.999% were melted in evacuated quartz ampoules at 850 °C for 16 h. Then the mixture allowed to cool down to 650°C very slowly at the rate of 3 °C per hrs. The mixture then kept at 650 °C for a day then cooled it down to room temp. The orientation of the $Bi_2Se_3$ single crystal was determined to be $Bi_2Se_3$(111) using x-ray diffraction (XRD) method. For STM studies the $Bi_2Se_3$ single crystal surface was prepared by cleaving the sample using clean room standard adhesive tape just before inserting it in the UHV sample preparation chamber. The surface was then annealed in situ at 670 K. The surface quality was checked with STM at 77 K. The images revealed atomic resolution of hexagonal arrangement of Se atoms, standard triangular defects, and atomic steps of 0.9 nm height that confirmed an atomically clean surface. FePc molecules were purchased from Sigma Aldrich and used as is for evaporation from a resistively heated quartz crucible attached to the UHV sample preparation chamber at 500 K onto the clean $Bi_2Se_3$ surface held at room temperature. STM imaging and spectroscopy (STS) were carried out at 77 K. Electrochemically etched tungsten tip was used for imaging. Several trials of imaging and spectroscopy were carried out with different tips. The images were processed using the image analysis software (SPIP 6.0.9, Image Metrology, Denmark). The local density of states (LDOS) was determined by scanning tunneling spectroscopy (STS). Using the standard lock-in technique a modulation voltage of 20 mV at 652 Hz frequency was applied with the SRS lockin amplifier. The tunneling voltage (V) was varied within ±2.5 V and the corresponding variation in the current (I) was obtained as a dI/dV spectrum, thus giving the tunneling conductance for the $FePc/Bi_2Se_3$ system.

RESULTS AND DISCUSSION

FePc molecules were deposited on a clean $Bi_2Se_3$ surface at room temperature. Figure 1a shows an STM image of FePc molecules after deposition on the $Bi_2Se_3$ surface. The molecules are seen as individual entities or small clusters of few molecules. The typical 4-lobe Pc structure, a characteristic of the square planar, $D_{4h}$ symmetry of the molecule can be easily identified. Also the center of the molecule is seen raised as a bright protrusion in the center. We refer to this configuration as phase A - raised 4-lobe. Thus the molecular recognition of FePc is straightforward and is similar to that of FePc on other substrates like Ag(110)[19] or Cu(111), Co(111), Au(111)[20-22]. The FePc molecules are scattered as individual molecules or small cluster-like structures on $Bi_2Se_3$. Molecular self-assembly has been studied in many systems[23]. FePc is known to exhibit large scale assembly on monolayer graphene on Ru(0001)[24] as well as a host template on the Kagome lattice of monolayer graphene on Ru(0001)[25,26]. On Au(110) FePc forms one dimensional (1D) nanochains[27]. But in our case, for the $Bi_2Se_3$ surface no ordered self-assembly was seen in several trials where the deposition parameters were varied. The FePc molecules appear diffuse. This has been verified at different imaging parameters and tips to rule out any tip effects. To improve the images, the $Bi_2Se_3$ surface was then annealed at 373 K and imaged again. After annealing the molecules appear sharp and very well resolved. Another striking observation is that in addition to phase A, the raised 4-lobe structure as defined earlier they can also be seen in a configuration where the center appears hollow, we call it phase B - the hollow 4-lobe structure, figure 1b.

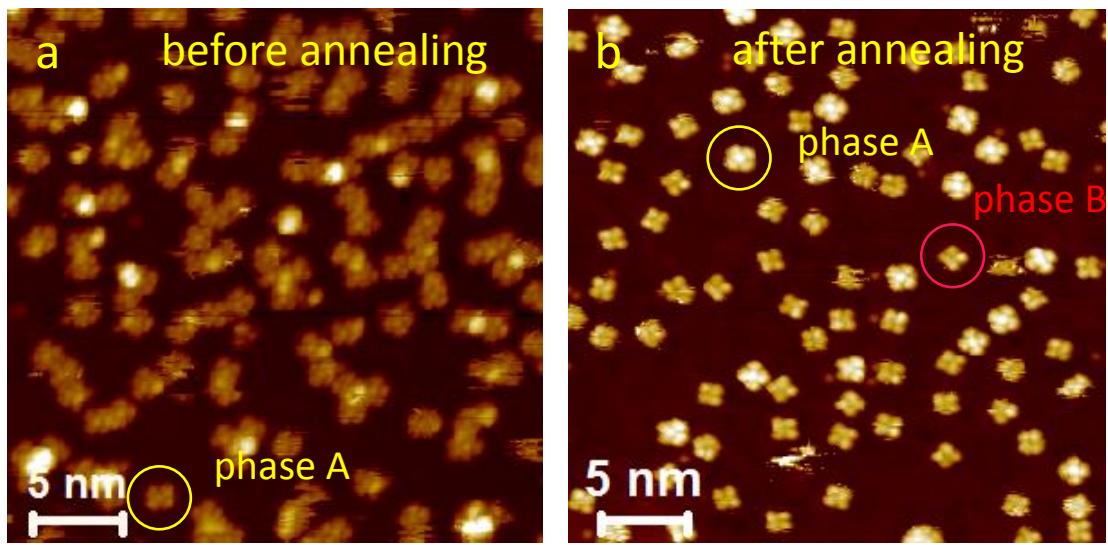

**Figure 1.** STM image of FePc molecules adsorbed on $Bi_2Se_3$ is shown in (a). The molecules show the typical 4 lobe structure of phthalocyanines. The center of the molecules is marked by a bright blob in the center. They appear diffuse. Only 1 phase of molecules is seen prior to annealing where the center of the molecule is visible as a bright

blob, referred to as phase A and marked by a circle in (a). Upon annealing the molecules are seen in 2 phases, phase A and phase B, marked by circles in (b). Phase B has a hollow center unlike phase A which has a bright blob in the center. For both images V = 2.2 V, I = 80 pA.

All the STM reports to date of FePc imaging have shown the phase A configuration irrespective of the substrate. Here the fact that we see phase B after annealing could be due to the energetic affinity of the different sites of the Se lattice for the Fe atom. Density functional theory (DFT) calculations on this system can help in a better understanding of this site selectivity. Atomic registry of the adsorbed molecules can be determined by using single atoms adsorbed on the surface as shown for the sexiphenyl molecules on Ag(111)[28] or by referring to the atomic resolution of the surface as in case of MnPc molecules on $Bi_2Te_3$ which is another TI surface[29] In our case, we acquired high resolution image of a single FePc molecule, both in phase A and in phase B, as shown in figure 2. More specifically, refer to figure 2(a). The FePc molecule in phase A can be seen in a high resolution scan with Se atoms in the background. To get the registry of the FePc molecule with respect to the Se lattice 3 dotted lines were drawn along the $[1\bar{1}0]$ directions of the Se atoms. The intersection of these lines points to the top site. The FePc molecule can be clearly seen adsorbed on this site, that is, on top of an Se atom. In figure 2(b) a close up STM image of FePc molecule in phase B is shown. Again, using the same method of registry, we see that in this phase B the FePc molecule is adsorbed on the bridge site, which is the location between 2 nearest neighbor Se atoms.

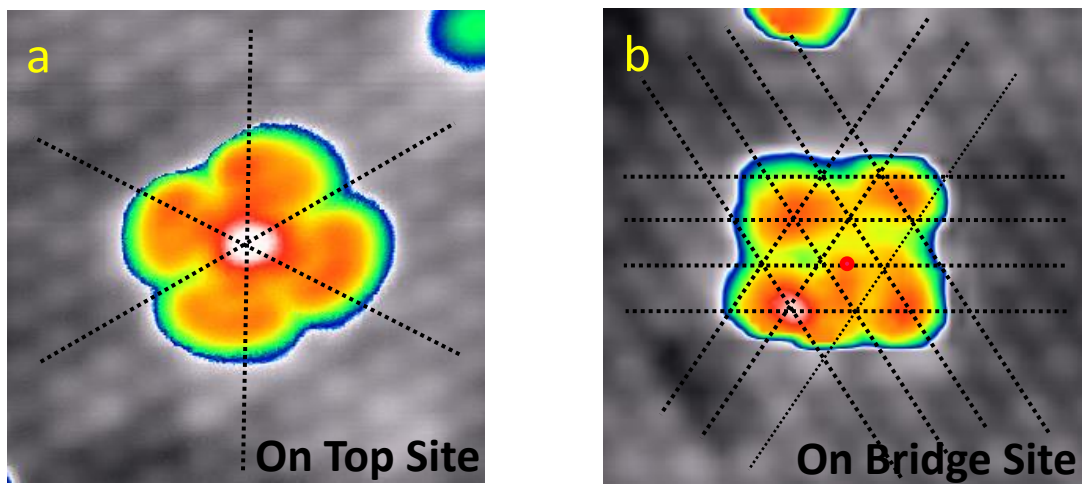

**Figure 2.** Single FePc molecule on $Bi_2Se_3$ in phase A configuration is shown in (a). The $Bi_2Se_3$ lattice is atomically resolved in the background of the molecule and the Se atoms can be seen. The dotted lines representing the $[1\bar{1}0]$ directions of the Se atomic arrangement are drawn as a guide to the eye to mark the positions of the atoms to ascertain the position of the FePc molecule with reference to the background atomic lattice. It can be clearly seen that the intersection of the dotted lines is an Se atom and the FePc molecule sits at this intersection, that is, it sits on the top site, or on top of an Se atom. Single FePc molecule on $Bi_2Se_3$ in phase B configuration is shown in (b). Using the same dotted line intersection method it can be seen that the FePc molecule sits on a bridge site, that is, in between 2 Se atoms.

STS measurements were performed on the bare $Bi_2Se_3$ substrate. From the dI/dV curve (black trace), figure 3, the $Bi_2Se_3$ states are observed at -1.2 eV and at 0.9 V. The $Bi_2Se_3$ states are likely to arise from the bulk $Bi_2Se_3$ states based on the calculated bulk band structure for $Bi_2Se_3$[30]. For single FePc molecules on the $Bi_2Se_3$ substrate, the STS measurements were done by positioning the tip at the center of the molecule. We observe that for FePc phase A the HOMO is at -1.1 V and LUMO is at 1.0 V whereas for FePc phase B the HOMO is at -1.0 V and LUMO is at 1.1 V. The HOMO and LUMO values of FePc are close to the reported values by other groups [31,32]. The small shift of the HOMO away from the Fermi level in phase A and the slight shift of the HOMO towards the Fermi level in phase B could be due to the site dependence of the FePc molecules. A theoretical insight from DFT calculations would be helpful to understand these shifts better.

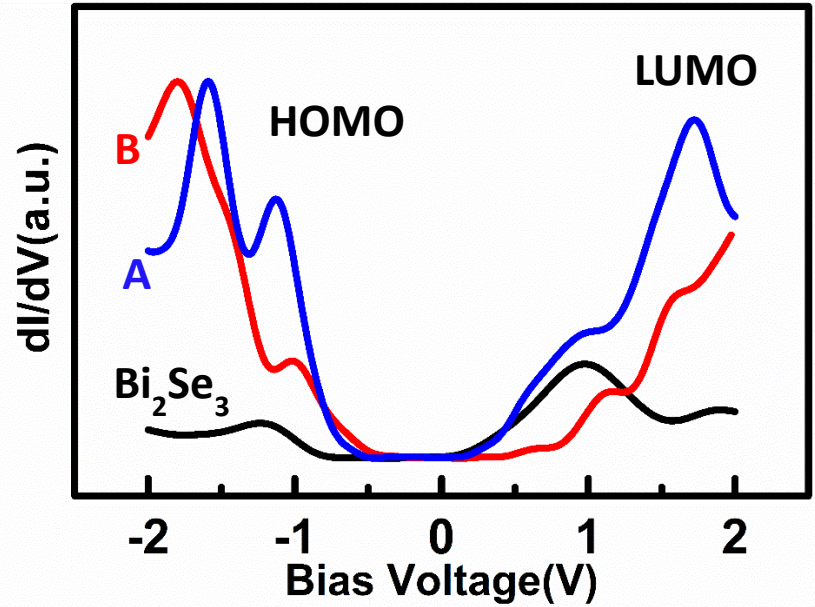

**Figure 3.** Point spectroscopy (STS) over bare $Bi_2Se_3$ surface is shown by the black trace, with states seen at -1.2 eV and 0.9 eV. The blue trace is the STS data for FePc molecule in phase A (residing on the top site of the $Bi_2Se_3$ surface), which pins the HOMO at -1.1 V and LUMO at 1.0 V. STS data for FePc molecule in phase B (residing on the bridge site of the $Bi_2Se_3$ surface) shows HOMO at -1.0 V and LUMO at 1.1 V.

The TISS of $Bi_2Se_3$, known to open a gap by Fe atom deposition [9], were then explored with STS measurements. These measurements were done close to the Fermi surface to see if the $Bi_2Se_3$ surface before doping with FePc molecules showed any change after the doping. An expected change is either a band gap opening as seen in some results of Fe on $Bi_2Se_3$[9] or a charge transfer process as seen in CoPc on $Bi_2Se_3$ [16]. A charge transfer resulting in the shift of the Dirac point was also seen in the case of MnPc on $Bi_2Te_3$[16,17,29].

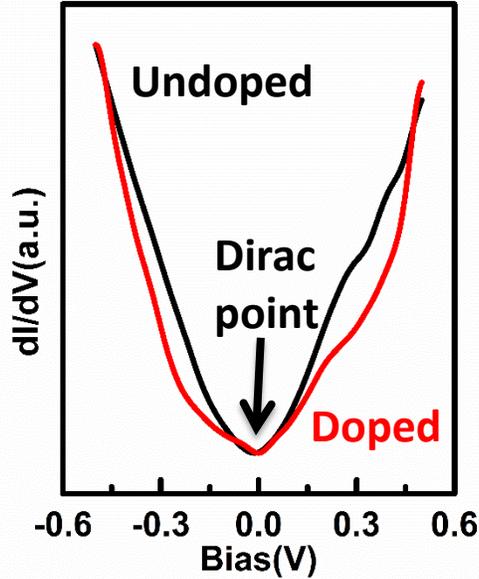

**Figure 4.** Point spectroscopy (STS) data for doped and undoped $Bi_2Se_3$ surface . The undoped $Bi_2Se_3$ surface shows the Dirac point at the Fermi energy (Ef), black/circle trace. The position of the Dirac point remains unchanged after the doping of the $Bi_2Se_3$ surface with FePc molecules, red/square trace.

For the FePc doped $Bi_2Se_3$ surface the STS data showed no change in the position of the Dirac point. The black curve in figure 4 shows the dI/dV spectrum taken on the bare $Bi_2Se_3$ surface, and the red curve shows the dI/dV curve taken on top of the FePc molecule. The Dirac point remains fixed at 0 eV. This indicates that there is no contribution from the magnetic Fe atom of FePc to perturb the topological surface state of $Bi_2Se_3$. Thus this magnetic phthalocyanine- TI interface keeps the surface state intact. Also there is no charge transfer between the molecule and the substrate unlike the MnPc/ $Bi_2Te_3$ system where the charge transfer that manifests as a shift of the Dirac point has been attributed to the band bending of $Bi_2Te_3$ caused by the molecule doping[29]. Comparing this with the ARPES study of CoPc on $Bi_2Se_3$[18] that shows charge transfer and 2H-Pc on $Bi_2Se_3$ (Supplementary information[18]) where the topological surface state is protected, we see that the central metal atom upon adsorption on the TI surface does influence its interaction with the exotic surface state of the TI.

CONCLUSIONS

To summarize we have investigated the magnetic phthalocyanine molecule FePc on the topological insulator surface $Bi_2Se_3$. Individual molecules of FePc are seen to adsorb on top site and hollow site of the $Bi_2Se_3$ surface, determined by concomitant background surface atomic resolution with the adsorbed single molecule resolution. No large scale ordering of FePc was observed on $Bi_2Se_3$. There is a slight shift of the HOMO and LUMO orbitals of FePc based on the adsorption site. The topological surface state of $Bi_2Se_3$ remains protected after the doping introduced by the FePc molecules. Our study suggests that the FePc- $Bi_2Se_3$ interface can be tailored for targeted spintronic applications favoring no backscattering. This study can be extended to other magnetic phthalocyanines and to magnetic porphyrins for exploring their effect on the topological insulator surface states.

AUTHOR INFORMATION
Corresponding Author
Email:aparna.d@iiserpune.ac.in

ACKNOWLEDGMENT
We gratefully acknowledge Dr. Surjeet Singh from IISER Pune for the crystals grown in his lab under DST project no. EMR/2016/003792.